\documentclass[aps,prl,twocolumn,groupedaddress,amsmath,amssymb,showpacs]{revtex4} 

\usepackage{graphicx}
\usepackage{dcolumn}
\usepackage{bm}

\begin{document}

\title{Manifestation of Topological Protection in Transport
Properties of Epitaxial Bi$_{2}$Se$_{3}$ Thin Films}

\author{A. A. Taskin}
\author{Satoshi Sasaki}
\author{Kouji Segawa}
\author{Yoichi Ando}
\email{y_ando@sanken.osaka-u.ac.jp}

\affiliation{Institute of Scientific and Industrial Research,
Osaka University, Ibaraki, Osaka 567-0047, Japan}

\date{\today}

\begin{abstract}

The massless Dirac fermions residing on the surface of three-dimensional
topological insulators are protected from backscattering and cannot be
localized by disorder, but such protection can be lifted in ultrathin
films when the three-dimensionality is lost. By measuring the
Shubnikov-de Haas oscillations in a series of high-quality
Bi$_{2}$Se$_{3}$ thin films, we revealed a systematic evolution of the
surface conductance as a function of thickness and found a striking
manifestation of the topological protection: The metallic surface
transport abruptly diminishes below the critical thickness of $\sim$6
nm, at which an energy gap opens in the surface state and the Dirac
fermions become massive. At the same time, the weak antilocalization
behavior is found to weaken in the gapped phase due to the loss of
$\pi$ Berry phase. 

\end{abstract}



\maketitle

In topological insulators (TIs) the energy states are fundamentally
modified from ordinary insulators by strong spin-orbit interactions,
giving rise to a topologically distinct state of matter with a gapped
insulating bulk and a gapless metallic surface \cite{H-K}. Various
interesting phenomena, including surface transport of spin-filtered
Dirac fermions that are immune to localization, have been predicted and
raised expectations for novel applications \cite{Moore,Q-SCZ,Fu-Kane}.
However, the progress in real applications of TIs crucially relies on
the ability to manipulate the surface current in transport experiments.
At present, such basic characterization as the surface conductance
measurement has been possible only in a few cases in single crystals
\cite{Ong_sci,BTS,Fisher_nm,BSTS,Morpurgo,Cd-BS,Cui,Sn-BTS} because of
the dominance of bulk transport caused by unintentional doping due to
defects. Molecular beam epitaxy (MBE) is a promising technique for the
synthesis of TIs \cite{Xue, Wu, Samarth, Xie, Oh, Wang} owing, in part,
to the relatively low deposition temperature at which defect
concentrations can be reduced from those in bulk crystals grown in
thermal equilibrium. So far, using MBE-grown films, angle-resolved
photoemission spectroscopy (ARPES) \cite{ARPES_MBE} and scanning
tunneling spectroscopy (STS) \cite{STM1} have provided useful
information about the topological surface state (SS), and in transport
experiments, such phenomena as weak antilocalization (WAL) and
gate-controlled ambipolar transport have been reported \cite{WAL1,WAL2}.
Also, since the MBE technique gives a precise control over film
thickness, transport measurements for widely varying surface-to-bulk
conductivity ratio have been performed \cite{Kim-Oh,e-e,Oh1}, although a
reliable separation of surface Dirac electrons from bulk carriers has
been hindered by a relatively low mobility of carriers in available thin
films. Recently, we have succeeded in growing high-quality epitaxial
films of Bi$_{2}$Se$_{3}$ that have a sufficiently high surface electron
mobility to present pronounced Shubnikov-de Haas (SdH) oscillations.
This made it possible to directly probe the surface conductance and the
topological protection of the SS. 

The immunity of the surface Dirac fermions to localization has a twofold
origin \cite{H-K,Moore,Q-SCZ}. One is the $\pi$ Berry phase associated
with massless Dirac fermions \cite{TAndo98}, which protects them from
weak localization through destructive interference of time-reversed
paths. The other is the peculiar spin-momentum locking which nulls the
backscattering probability \cite{H-K,Moore,Q-SCZ}. Those mechanisms are
collectively called topological protection. Recently, it was found
\cite{ARPES_MBE,Linder,OscCros,UltraThin,VirginiaTech2010,Austin2011}
that when TIs are thinned to the extent that the top and bottom surface
states feel each other, their hybridization leads to an opening of the
gap at the Dirac point and results in a degenerate, massive Dirac
dispersion. This gapped state obviously violates the topological
protection, but its consequence in the surface transport properties has
not been duly addressed. In fact, this question is important because
recently a lot of attention has been paid to the way to open a gap at
the Dirac point \cite{Sato} to realize topological magnetoelectric
effects \cite{Qi}. One may expect that unless the chemical potential is
located exactly within the gap, the metallic surface transport is
largely unaffected by the gap opening because states are kept being
available for transport at the Fermi level. However, in the present work
it turned out that the change in the Dirac spectrum deep in the occupied
state has a profound effect on the physics at the Fermi level. 

The growth of Bi$_{2}$Se$_{3}$ films occurs in a layer-by-layer manner,
in which the 0.95-nm thick Se-Bi-Se-Bi-Se quintuple layer (QL)
constitutes the basic unit \cite{ARPES_MBE}. Our systematic
magnetotransport measurements for varying thickness reveal a sudden
diminishment of the surface transport below the critical thickness of
$\sim$6 QL, below which the energy gap opens in the Dirac spectrum
\cite{ARPES_MBE}. We also observed that the weak antilocalization
behavior \cite{WAL1,WAL2} quickly weakens below the critical thickness.
We discuss that those striking effects are due to acquired degeneracy of
the surface states \cite{UltraThin} and loss of their $\pi$ Berry phase
\cite{SQS1,SQS2} in the gapped phase.

Our MBE films were grown under Se-rich conditions on insulating sapphire
(0001) substrates whose size was approximately 15 $\times$ 4 mm$^{2}$.
To obtain films of high enough quality to present SdH oscillations, we
employed a two-step deposition procedure \cite{footnote0, SM}. Both Bi
(99.9999\%) and Se (99.999\%) were evaporated from standard Knudsen
cells. The Se$_{2}$(Se$_{4}$)/Bi flux ratio was kept between 15 -- 20.
The growth rate, which is determined by the Bi flux, was kept at 0.2 --
0.3 QL/min. The resistance $R_{xx}$ and the Hall resistance $R_{yx}$
were measured in the Hall bar geometry by a standard six-probe method on
rectangular samples on which the contacts were made with silver paste at
the perimeter and cured at room temperature under pressure of $\sim$1
Pa. The magnetic field was swept between $\pm$ 14 T at fixed
temperatures and was always applied perpendicular to the films, except
for the angular-dependence measurements of the SdH oscillations. 

\begin{figure}\includegraphics*[width=8.5cm]{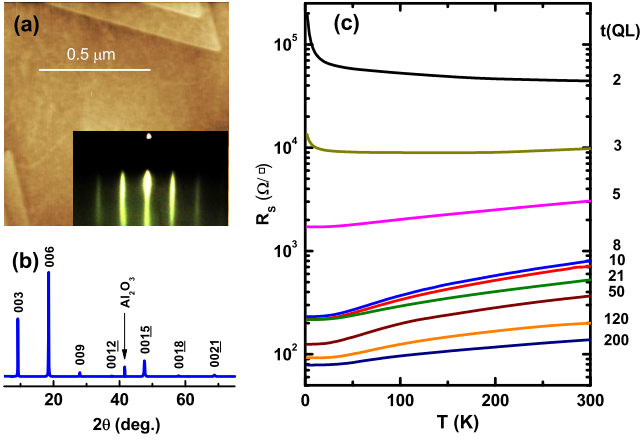}
\caption{(Color online) 
(a) AFM image of a 50-nm thick Bi$_{2}$Se$_{3}$ film showing atomically 
flat terraces with 1-QL steps. Inset: typical RHEED pattern.
(b) X-ray diffraction pattern of a 200-QL film.
(c) Temperature dependences of $R_{s}$ for different thickness.
}
\label{fig1}
\end{figure}

An atomic force microscopy (AFM) image of our relatively thick (50 QL)
film is shown in Fig. 1(a), where a large, atomically flat area of $\sim$1
$\mu$m$^{2}$ and several sharp terraces with the height of exactly 1 QL
can be clearly seen. The reflection high-energy electron diffraction
(RHEED) pattern with sharp 1$\times$1 streaks [Fig. 1(a) inset] and sharp
x-ray diffraction peaks [Fig. 1(b)] are also indications of the high
crystal quality of our films.

Temperature dependencies of the sheet resistance, $R_{S}(T)$, measured in
films with systematically changed thickness $t$ down to 2 QL are shown
in Fig. 1(c). The $R_{S}(T)$ behavior is metallic in thick films, but
below $t$ = 5 QL, it starts to show an upturn at low temperatures. In
particular, the sharp divergence in $R_{S}$ for $T \rightarrow 0$ in the
2-QL film is indicative of strong Anderson localization and an
insulating ground state \cite{SM}.

\begin{figure}\includegraphics*[width=8.5cm]{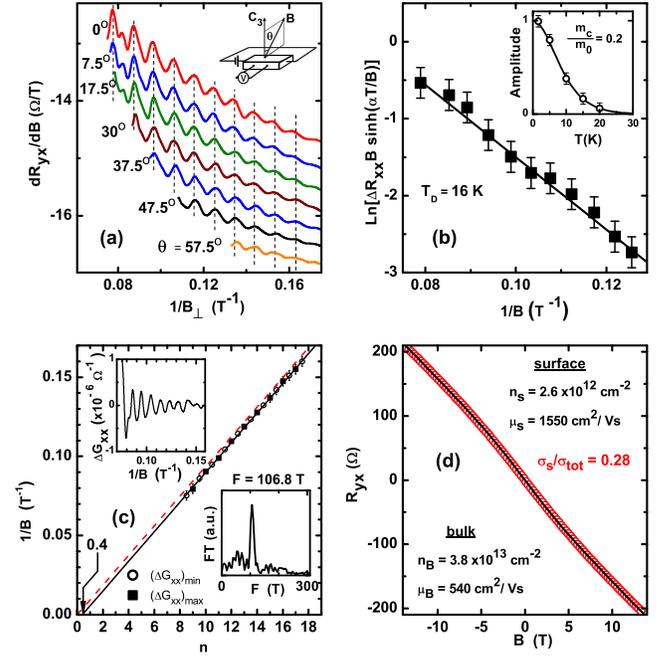}
\caption{(Color online) 
Surface SdH oscillations in the 10-QL film. 
(a) $dR_{yx}/dB$ in tilted magnetic fields, plotted 
as a function of $1/B_{\perp}$ ($=1/B\cos\theta$); curves 
are shifted vertically for clarity.
Dashed lines mark the positions of maxima.
Inset shows the geometry of the experiment.
(b) Dingle plot of the oscillations in $\Delta R_{xx}$ at 1.6 K,
obtained after subtracting a smooth background from $R_{xx}(B)$, giving
the Dingle temperature of 16 K. Inset: $T$-dependence of the SdH
amplitude for $\theta$ = 0$^{\circ}$. 
(c) Landau-level fan diagram for oscillations in $G_{xx}$ measured at
$T$ = 1.6 K and $\theta $ = 0$^{\circ}$; following Ref.
\cite{Ong_arxiv}, integers $n$ (half-integers $n+\frac{1}{2}$) are
assigned to the minima (maxima) in $\Delta G_{xx}$. The solid line is a
linear fitting to the data with the slope fixed at $F$ = 106.8 T;
the dashed line has the same slope and extrapolates to zero. 
Upper inset shows $\Delta G_{xx}$
vs. 1/$B$ after subtracting a smooth background; lower inset shows its
Fourier transform giving $F$ = 106.8 T. (d)
Fitting of the two-band model to the $R_{yx}(B)$ data at 1.6 K.
}
\label{fig2}
\end{figure}

The breakthrough in the present work is that our films exhibit
pronounced two-dimensional (2D) SdH oscillations to provide a direct way
to probe the surface charge transport. As an example, the analysis of
the SdH oscillations in the 10-QL film is shown in Fig. 2. The 2D
character of the oscillations is evident in Fig. 2(a), where the
positions of the maxima and minima depend only on the perpendicular
component of the magnetic field, $B_{\perp}$. The oscillation frequency
$F$ = 106.8 T is obtained from the Fourier transform [lower inset of
Fig. 2(c)], and this is a direct measure of the Fermi wave number $k_F$
= 5.7 $\times$ 10$^6$ cm$^{-1}$. As we discuss in detail in the
Supplemental Material
\cite{SM}, if the SdH oscillations are due to the trivial 2D electron
gas (2DEG) which may form due to a band bending near the surface
\cite{Hofmann}, this $k_F$ is so small that it imposes too strong a
constraint on the possible bulk Fermi level, which makes it impossible
to consistently explain the transport data. Hence, we identify the
oscillations to be due to surface Dirac fermions, and the obtained $k_F$
gives their density $n_{s}$ = 2.6 $\times$ 10$^{12}$ cm$^{-2}$.

In this 10-QL film, we observed only a single frequency, but we usually
see two frequencies in other films (see Fig. S3 of the Supplemental Material \cite{SM}),
suggesting that the top and bottom surfaces have somewhat different
$n_{s}$. The temperature dependence of the oscillation amplitude [Fig.
2(b) inset] gives the cyclotron mass $m_{c}$ = 0.2$m_{0}$ ($m_0$ is the
free electron mass) \cite{Shoenberg} which in turn gives the Fermi
velocity $v_{F}$ = 3.3 $\times$ 10$^{7}$ cm$\,$s$^{-1}$. This $v_{F}$ is
consistent with the ARPES data \cite{Hasan2009} as well as the STS data
\cite{STM1,STM2} for the Dirac cone. The obtained $k_F$ value
corresponds to the Fermi level of $\sim$0.16 eV above the Dirac point
for the topological surface state, which points to a slight upward band
bending \cite{footnote1}. The Dingle analysis [Fig. 2(b), see the Supplemental Material
\cite{SM} for details] yields the mobility $\mu_{s}$ = 1330
cm$^2$V$^{-1}$s$^{-1}$. Finally, Fig. 2(c) shows the Landau-level fan
diagram for the oscillations in conductance $G_{xx}$, where the
positions of the minima in $\Delta G_{xx}$ (shown in the upper inset)
are plotted as a function of $n$ \cite{Ong_arxiv}. Here, to minimize the error
occurring from extrapolation \cite{Sn-BTS}, we fix the slope of the
linear fitting by using $F$ = 106.8 T obtained from the Fourier analysis
and determine the intercept $n$ = 0.40 $\pm$0.04 (solid line); this is
very close to the ideal value of 0.5 for Dirac electrons bearing the
$\pi$ Berry phase \cite{BerryPh}, giving further confidence in the
origin of the SdH oscillations. For comparison, a straight line with the
same slope to give zero Berry phase is shown in Fig. 2(c) with a dashed
line, which is obviously inconsistent with the experimental data.

To estimate the contribution of the SS in the overall transport in this
10-QL film, we use the magnetic-field dependence of the Hall
resistivity, $R_{yx}(B)$ [Fig. 2(d)], which is not linear in $B$ and
thus signifies the presence of at least two types of carriers. The
fitting of a standard two-band model \cite{BTS,BSTS,A-M} to the data, in
which $n_s$ is fixed by the SdH frequency, gives the surface
contribution to the total conductance, $G_{s}/G_{tot}$, of 28\%. The
$\mu_{s}$ value obtained from this fitting is close to the SdH result,
assuring the consistency of our analysis.

\begin{figure}\includegraphics*[width=8.5cm]{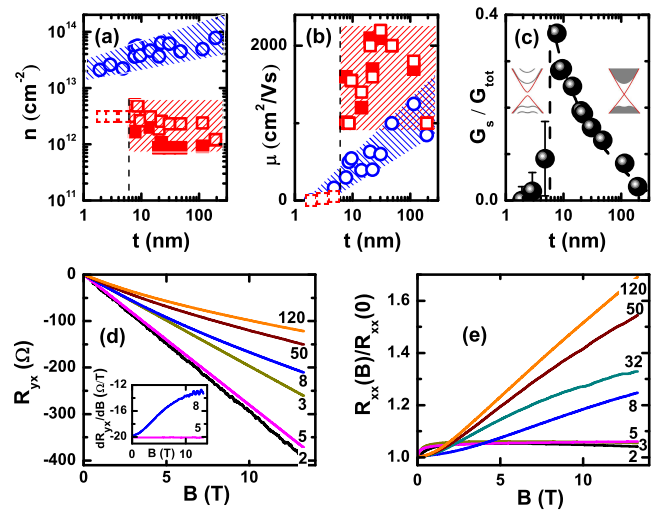}
\caption{(Color online) 
(a) Squares are the $n_s$ obtained from SdH oscillations (open and
filled squares represent two different surfaces of the same film)
and the dashed squares are an extrapolation of the trend above $t_c$;
circles are the sheet density of bulk carriers obtained from two-band
analyses. 
(b) Mobilities of surface (squares) and bulk (circles) carriers obtained 
from two-band analyses.
(c) $G_{s}/G_{tot}$ obtained from the data in (a) and (b). 
Insets are schematic pictures of the energy bands for the two regimes.
(d) $R_{yx}(B)$ at 1.6 K for various thickness shown in QL unit; 
inset shows the derivative $dR_{yx}/dB$
for the 5 and 8 QL films (5-QL data are shifted for clarity).
(e) $R_{xx}(B)/R_{xx}(0)$ of the same films.
}
\label{fig3}
\end{figure}

The same analysis can be applied to all measured films with $t \ge$ 8
QL, in which we consistently observed SdH oscillations \cite{SM}.
Evolutions of the transport parameters with changing $t$ are summarized
in Figs. 3(a)-3(c). We note that by tracing the evolution of the SdH
oscillations starting from thick films, we can distinguish the
topological SS from the 2D quantum-well (QW) state of the bulk
origin \cite{SM} as the source of the SdH oscillations.

Our main finding is that the surface transport abruptly diminishes below
a critical thickness $t_{c}$ which is located between 5 and 8 QL. This
change is most convincingly manifested in the behavior of $R_{yx}(B)$,
which suddenly becomes $B$-linear in films with $t \le$ 5 QL [Fig.
3(d)]; this indicates that the transport becomes suddenly dominated by
only one type of carriers. Correspondingly, the SdH oscillations
disappear for $t < t_c$. More quantitatively, assuming that $n_s$ is
essentially unchanged through $t_c$ [dashed squares in Fig. 3(a)], 
one can estimate that $\mu_s$ must be suddenly degraded
by more than an order of magnitude in samples with $t < t_c$
[dashed squares in Fig. 3(b)] for $R_{yx}(B)$ to become linear and be
governed by bulk carriers \cite{SM}. The magnetoresistance behavior
shown in Fig. 3(e) also presents a qualitative change below $t_{c}$,
showing a negative slope at high fields. This evolution is best
represented in the $t$ dependence of $G_{s}/G_{tot}$ [Fig. 3(c)], which
shows a steady increase with decreasing $t$ to reflect the change in the
surface-to-bulk ratio, but it drops sharply below $t_{c}$ to signify
that the surface transport is abruptly diminished. 

This observation naturally calls for the question whether the observed
diminishment of the surface transport in ultrathin films might be
related to a lowering of the quality in thinner
films. In this respect, our ultrathin films remain essentially flat and
smooth across $t_c$, with the surface bumpiness of only $\sim$1 QL 
(see Fig. S10 of the Supplemental Material \cite{SM}). This observation, combined with the fact that
$G_{s}/G_{tot}$ increases steadily with decreasing $t$ until it reaches
$t_c$, testifies against the above concern.

\begin{figure}\includegraphics*[width=8.5cm]{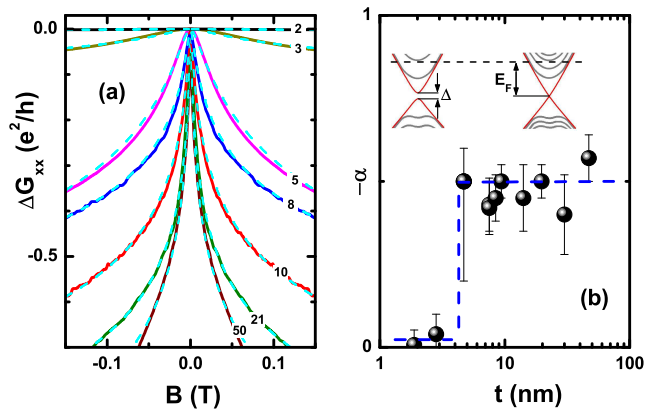}
\caption{(Color online) 
(a) WAL behavior in  sheet conductance at 1.6 K for various 
thicknesses shown in QL units; dashed lines are the fittings using Eq. (1). 
(b) Thickness dependence of $\alpha$. 
Inset shows schematic energy bands above and below the critical thickness.
}
\label{fig4}
\end{figure}

In addition to the above results, we found a striking change in
the WAL behavior \cite{WAL1,WAL2} below $t$ = 5 QL. 
Figure 4(a) shows the magnetoconductance of our films measured in
perpendicular magnetic fields at 1.6 K. Dashed lines are the fitting
with the Hikami-Larkin-Nagaoka formula \cite{H-L-N},
\begin{equation}
\Delta G_{xx}(B) = \alpha
\frac{e^{2}}{\pi h}    \left [  \Psi \left( \frac{\hbar}{4 e L_{\phi}^2 B} + 
\frac{1}{2} \right) 
- {\rm ln} \left( \frac{\hbar}{4 e L_{\phi}^2 B} \right)  \right ],
\end{equation}
where $\Psi$ is the digamma function and $L_{\phi}$ is the phase
coherence length. The prefactor $\alpha$ should be $-\frac{1}{2}$ for
each transport channel that either carries a $\pi$ Berry phase
\cite{TAndo98} or bears a strong spin-orbit interaction \cite{H-L-N}. In
our analysis, $\alpha$ and $L_{\phi}$ are the only fitting parameters
and Fig. 4(b) shows the $t$ dependence of obtained $\alpha$ (see the Supplemental Material
\cite{SM} for details). For $t \ge$ 5 QL, we observed $\alpha \approx
-\frac{1}{2}$ similar to that reported for metallic Bi$_{2}$Se$_{3}$
thin films where top and bottom surfaces are connected though bulk
electrons \cite{WAL1,WAL2}. The change to $\alpha \approx 0$ observed
for $t \le$ 3 QL is in accord with the diminishment of the surface
transport channel and the eventual localization of the bulk state. A
similar tendency was also observed in previous 
studies \cite{Kim-Oh,e-e,Oh1}, although the decrease in $\alpha$ was less
pronounced probably due to a larger metallicity of the measured samples. 

Now we discuss the origin of our observations. As was already found in
Bi$_{2}$Se$_{3}$ ultrathin films by ARPES \cite{ARPES_MBE}, an energy
gap in the SS opens at the Dirac point below $t_c \simeq$ 6 QL.
Obviously, our transport measurements reflect this change in the Dirac
dispersion. The gap is due to hybridization between top and bottom
surfaces [schematically shown in the Fig. 4(b) inset], and such a hybridization
gap $\Delta$ changes the massless Dirac dispersion $E = \pm\hbar
v_{F}k$ into a massive one $E = \pm \sqrt{(\hbar v_{F} k)^2 + (\Delta/2)^2}$
\cite{UltraThin,SQS1}. For Bi$_{2}$Se$_{3}$, $\Delta$ is about 0.25 eV
at $t$ = 2 QL \cite{ARPES_MBE}. In our films, the Fermi level is
estimated to be $\sim$0.16 eV above the Dirac point, so
even in our thinnest film the Fermi level is not in the gap but crosses
the SS as schematically shown in the inset of Fig. 4(b). Hence, the
observed drastic suppression of the surface transport is {\it not} due
to the disappearance of surface carriers but is likely due to an
enhanced scattering of the carriers \cite{note3}.

The gap opening also has a profound effect on the Berry phase
$\phi_{B}$ of the surface band $\psi_{k}(r)$. In the simplest case, 
$\phi_{B}$ is given by \cite{SQS1}
\begin{equation}
\phi_{B} = -i \int_{0}^{2\pi} d\varphi \left\langle \psi_{k}(r) \Bigg 
\vert \frac{\partial \psi_{k}(r)}{\partial \varphi} \right\rangle= 
\pi \left( 1- \frac{\Delta}{E_{F}} \right) ,
\end{equation}
and hence the Berry phase is reduced from $\pi$ when a gap opens. This
lifts the immunity of the SS from weak localization. Also, this change
in $\phi_{B}$ weakens the WAL in the SS \cite{SQS2}. It is useful to
note that in Fig. 4(a) the WAL was still observed for $t$ = 5 QL which
is below $t_c$, but this is natural because the small $\Delta$ at this
$t$ \cite{ARPES_MBE} makes the deviation of $\phi_{B}$ from $\pi$ to be
small. 

More importantly, in the gapped phase, the SS becomes degenerate
\cite{UltraThin}, which means that now for each momentum both up and
down spin states are available. This opens the backscattering channel
and significantly reduces the surface mobility. Actually, the
hybridization of top and bottom surfaces means that the system is no
longer truly three-dimensional (3D), so it is natural that the
topological properties of 3D TIs are lost in the gapped phase, in full
agreement with recent {\it ab initio} density functional studies of TI
thin films \cite{VirginiaTech2010,Austin2011}. In this respect, the
present observation is a spectacular manifestation of the topological
protection of the SS in 3D TIs.

\begin{acknowledgments} 

We thank V. G. Mansurov for helpful suggestions about MBE growth. 
We also thank S. Oh, M. Sato, and Y. Tanaka for useful discussions.
This work was supported by JSPS (NEXT Program), MEXT (Innovative Area 
``Topological Quantum Phenomena" KAKENHI), and AFOSR (AOARD 104103 and
124038).

\end{acknowledgments}

\clearpage
\widetext

\renewcommand{\thefigure}{S\arabic{figure}} 

\setcounter{figure}{0}

\renewcommand{\thesection}{S\arabic{section}.} 

\begin{flushleft} 
{\Large {\bf Supplemental Material}}
\end{flushleft} 

\vspace{2mm}

\begin{flushleft} 
{\bf S1. Van der Waals epitaxy and two-step deposition}
\end{flushleft} 

The building block of the layered Bi$_{2}$Se$_{3}$ is a quintuple layer
(QL) as depicted in Fig. S1(a). The surface is terminated by Se atoms
without dangling bonds, interacting only by weak van der Waals forces.
Due to the weak interaction with a substrate, an epitaxial growth with
its own lattice constant is possible at the very beginning (with a
drastically relaxed lattice matching condition) as schematically shown
in Fig. S1(b). Thus, the growth of ultrathin films starting from one QL is
possible \cite{Koma}. This is called van der Waals epitaxy.

One of the key parameters of the MBE growth is the temperature of
deposition, which determines the rate of adsorption/desorption of Bi and
Se as well as their surface diffusion. We found that increasing the
substrate temperature greatly improves the size of grown terraces.
Unfortunately, at high temperatures, the growth of thin films is
virtually impossible due to a weak interaction of adatoms with the
substrate. The crucial improvement was achieved by using a two-step
deposition method \cite{footnote0}: In the first step, the initial 1 QL
of Bi$_{2}$Se$_{3}$ is deposited at a low temperature of
110--120$^\circ$C; then, the temperature of the substrate is slowly
raised up to 300--320$^\circ$C, and, in the second step, the desired
number of QLs is deposited. We obtained the best quality thin films at
the temperature of the second step of around 300$^\circ$C.

\begin{flushleft} 
{\bf S2. Anderson localization}
\end{flushleft} 

An upturn in the sheet resistance in ultrathin 
Bi$_{2}$Se$_{3}$ films has always been seen at low temperatures 
\cite{e-e,Kim-Oh,Oh1}, in spite of the difference in concentration of charge 
carriers and their mobilities. Figure S2 shows the diverging
trend of $R_{s}$ at low temperatures we observed in the 2-QL film. A
rapid decrease in the mobility with decreasing temperature is also shown
in the inset of Fig. S2, which points to a localization of charge carriers in
the zero-temperature limit. 
Similar behavior of $R_{s}(T)$ was studied in Ref. \cite{e-e} on films with 
a higher conductance, where the insulating ground state was proposed to be
due to strong interactions of surface electrons, which were assumed to
dominate the transport in ultrathin films. 
However, for ultrathin films with lower conductance, which is determined by
the concentration of charge carriers and the strength of disorder, 
the Anderson localization of a disordered 2D metallic conductor
seems to be a better (more natural) explanation for the observed insulating
state. Such localization behavior in low-conductance ultrathin films highlights 
the disappearance of topological protection of the surface state.
This conclusion is also supported by a recent study of
field-effect devices fabricated from ultrathin ($\sim$3.5 nm)
Bi$_{2}$Se$_{3}$ crystals obtained by mechanical exfoliation on
SiO$_{2}$/Si substrates \cite{FuhrerNano}. By measuring the temperature
and gate-voltage dependences of the conductance, the authors of Ref.
\cite{FuhrerNano} observed clear insulating behavior with an activated
energy gap when the chemical potential is tuned into the gap, which made
them conclude that their 3.5-nm thick Bi$_{2}$Se$_{3}$ crystals are
conventional insulators.

\begin{flushleft} 
{\bf S3. SdH oscillations}
\end{flushleft} 

Transport properties of surface carriers are obtained mostly form the
SdH oscillations observed in films with the thickness $t$ above 6 QL
(Fig. S3). The frequency $F$ identified in SdH oscillations in $R$ vs.
$1/B$ gives the size of a 2D Fermi surface (FS) via the Onsager relation
$F = (\hbar / 2\pi e)A$, where $A=\pi k_{F}^{2}$ is the FS cross
section; $\hbar$ and $e$ are the reduced Plank's constant and 
the electron charge, respectively; $k_{F}$ is the
Fermi wave number. The concentration of spin-filtered surface charge
carriers is $n_s$ = $k_{F}^{2}/4 \pi$.

The cyclotron mass $m_{c}$ is obtained from the temperature damping
factor of the first harmonic, $R_{T} = \pi \lambda/\sinh(\pi \lambda)$,
where $\lambda = 2\pi k_{B}T/\hbar \omega_c$ and $\omega_c \equiv
eB/m_{c}$ is the cyclotron frequency \cite{Shoenberg}. From $k_F$ and
$m_{c}$, we obtain the Fermi velocity $v_{F} = \hbar k_{F}/m_{c}$. 

The quantum scattering time $\tau = \hbar / (2 \pi k_{B} T_{D})$ and the
mean free path $\ell_{s}$ = $v_{F} \tau$ are obtained from the Dingle
damping factor $R_{D} = \exp(-2\pi^{2} k_{B}T_{D} / \hbar \omega_c)$,
which accounts for scattering effects \cite{Shoenberg}; the surface
mobility $\mu_s$ is estimated from the relation for the surface
conductance $G_{s} = e \mu_{s} n_{s} = (e^{2}/h) (k_{F}\ell_{s})$.

\begin{flushleft} 
{\bf S4. Sudden drop in the surface mobility below $t_c$}
\end{flushleft} 

As has been discussed in the main text, the SdH oscillations disappear
below $t_c$, which coincides with the disappearance of topological
protection of the surface state (SS) when the energy gap opens and the
Dirac fermions become massive. Here we present additional discussions
that the disappearance of the SdH oscillations with decreasing film
thickness cannot be simply due to a smooth decrease of the surface
mobility below the observational limit.

Figures S4 and S5 show $R_s(B)$ and $R_{yx}(B)$ data measured in films
with $t$ = 8 and 5 QL, respectively, together with their fittings
considering the surface and bulk channels, similar to what has been
described in the main text for $t$ = 10 QL [Fig. 2(d)]. For the 8-QL
film (Fig. S4), where we observed two SdH frequencies (Fig. S3), the
concentrations of surface carriers $n_{1}$ = 1.7 $\times$ 10$^{12}$
cm$^{-2}$ and $n_{2}$ = 5.0 $\times$ 10$^{12}$ cm$^{-2}$ are fixed by
the SdH frequencies. For the 5-QL film (Fig. S5), where SdH oscillations
were not detected, we assume $n_1$ = $n_2$ = 3 $\times$ 10$^{12}$
cm$^{-2}$ following the trend in $n_s$ vs $t$ for the surface channel
[see Fig. 3(a)].
 
A three-band model analysis for the 8-QL film gives the mobilities
$\mu_{s1}$ = 1600 cm$^{2}$V$^{-1}$s$^{-1}$ and $\mu_{s2}$ = 1000
cm$^{2}$V$^{-1}$s$^{-1}$ for two different surfaces and $\mu_3$ = 161
cm$^{2}$V$^{-1}$s$^{-1}$ for the bulk carriers (Fig. S4). Note that this
combination of carrier densities and mobilities for the three channels
reproduce the curved $R_{yx}(B)$ behavior as well as the 
$B$ = 0 T value of $R_s$, but the observed $B$ dependence
of $R_s$ is stronger than the three-band model prediction; this is
understandable, because the simple multi-band model employed here
neglects the intrinsic magnetoresistance ({\it i.e.} the $B$ dependence
of the mobility) which is known to exist in Bi$_2$Se$_3$.
 
In contrast, the $R_{yx}$ data for the 5-QL film is almost perfectly
linear in $B$, and to reproduce this linear behavior while keeping the
consistency with the $R_s(0)$ value, one needs to consider a very low
mobility for the surface channel. For example, a combination of the
surface mobility $\mu_{1,2}$ = 50 cm$^{2}$V$^{-1}$s$^{-1}$ and the bulk
mobility $\mu_3$ = 140 cm$^{2}$V$^{-1}$s$^{-1}$ gives reasonable
reproductions of $R_s(0)$ and $R_{yx}(B)$, as shown in Fig. S5. On the
other hand, if one assumes a gradual change in the surface mobility
across $t_c$ and hypothetically put $\mu_{1,2}$ = 500
cm$^{2}$V$^{-1}$s$^{-1}$, it is impossible to make $R_s(0)$ to be
consistent with the data nor to make $R_{yx}(B)$ linear, as shown in
Fig. S6. Therefore, one can conclude that the surface mobility indeed
dropped by more than one order of magnitude when $t$ was reduced from 8
to 5 QL.

\begin{flushleft} 
{\bf S5. Quantum confinement effects}
\end{flushleft} 

\begin{flushleft} 
{\bf A. Formation of quantum-well states in ultrathin films}
\end{flushleft} 
 
\vspace{-4 mm}

In thin films, the effects of quantum confinement modify the 3D bulk
energy states to 2D quantum-well (QW) states when the film thickness
becomes comparable to the de Broglie wavelength of the carriers. In the
case of very thin films, it is usually difficult to distinguish between
a genuine SS and a 2D QW state if one simply measures the angular
dependence of the SdH oscillations. Fortunately, we were able to measure
SdH oscillations in all films with $t >$ 6 QL, including those in which
bulk states are certainly in the 3D limit. As an example, Fig. S7 shows
the 2D character of SdH oscillations measured in a 120-QL film. It is
worth noting that in none of the measured films we saw SdH oscillations
coming from the 3D bulk states. This is most likely because of
unfavorable conditions for their observations, that is, the relatively
low mobility and a high bulk-carrier concentration together make the
oscillation amplitude to become small \cite{Adams}. The systematic
measurements of SdH oscillations in a wide range of thickness make it
possible to reliably distinguish the SSs from bulk states or the 2D QW
states by tracing them from thick samples.

\begin{flushleft} 
{\bf B. Surface band bending and resulting 2D electron gas}
\end{flushleft} 

\vspace{-4 mm}

The surface bend bending can create a 2D electron gas (2DEG) on the
surface of TIs, and this could be responsible for the observed 2D SdH
oscillations instead of the topological SSs. Spectroscopically,
coexistence of the topological state and a 2DEG state on the surface of
Bi$_{2}$Se$_{3}$ has been observed by ARPES a few hours after cleaving
\cite{Hofmann,King}, which is believed to be due to gradual formation of
defects (Se vacancies) on a freshly cleaved surface in ultrahigh vacuum.
Figure S8 reproduces the energy dispersions of the SS and the 2DEG state
measured along the $\bar{K}-\bar{\Gamma}-\bar{K}$ direction in Ref.
\cite{Hofmann}. Solid lines are our fittings of the equations shown in
Fig. S8 to the experimental data.

Now we show that it is not possible to ascribe the observed 2D SdH
oscillations to the 2DEG state. The frequency of the SdH oscillations in
the 10-QL film shown in Fig. 2 of the main text corresponds to $k_{F}$ =
5.7 $\times$ 10$^{6}$ cm$^{-1}$. As discussed in the main text, if the
SdH oscillations are coming from the topological SS, this $k_{F}$ gives
the estimate for $E_F$ of 160 meV measured from the Dirac point. On the
other hand, if this $k_{F}$ is associated with the 2DEG state shown in
Fig. S8, $E_F$ should be 82 meV from the bottom of this 2DEG state and
the corresponding density of the 2DEG, $n_{\rm 2DEG}$, is 5.2 $\times$
10$^{12}$ cm$^{-2}$. One should remember that for this 2DEG state to
form, the conduction band must be bent {\it downward} near the surface,
which dictates that the $E_F$ of the bulk state measured from the bulk
conduction band edge must be {\it smaller} than 82 meV. This means that
the maximum possible bulk carrier density is 5 $\times$ 10$^{18}$
cm$^{-3}$ if we use the effective mass $m^*$ = 0.13$m_0$ for the bulk
conduction band \cite{Analytis}. (A larger bulk $E_F$ would mean that
the band is bent {\it upward} near the surface, which precludes the
formation of the 2DEG state from the beginning.)

However, the high-field slope of the $R_{yx}(B)$ data, which simply
gives a sum of the densities of electrons from all bands \cite{A-M},
points to the existence of a total of 4 $\times$ 10$^{19}$ cm$^{-3}$ of
electrons in our 10-QL film; since the above 2DEG density corresponds to
the volume density of only 5.2 $\times$ 10$^{18}$ cm$^{-3}$, the maximum
possible bulk carrier density of 5 $\times$ 10$^{18}$ cm$^{-3}$ is
not enough to account for the Hall data. [Note that the density of Dirac
electrons is 1 $\times$ 10$^{13}$ cm$^{-2}$ at most in the above 2DEG
scenario (see later), and hence the topological SS can only provide up
to 2 $\times$ 10$^{19}$ cm$^{-3}$ of carriers from the two surfaces, and
this is still not enough.] Therefore, one cannot achieve a consistent
understanding of the transport data in the 2DEG scenario.

To illustrate the difficulty of the 2DEG scenario in a more quantitative
manner, let us assume, for simplicity, the same band bending as observed
in Ref. \cite{Hofmann} ({\it i.e.} the main 2DEG state located 40 meV
below the bottom of the bulk conduction band and 90 meV above the
conduction-band edge of the surface, which gives the total of 130 meV
downward band bending, see the inset of the left panel of Fig. S9)
\cite{Poisson}. Then the position of the Dirac point (DP) is about 370
meV below the surface Fermi level, giving the concentration of Dirac
electrons $n_{\rm Dirac}$ = 1 $\times$ 10$^{13}$ cm$^{-2}$. The bulk
$E_F$ should be 42 meV in this situation, and the corresponding bulk
carrier density is $n_{\rm bulk}$ = 1.8 $\times$ 10$^{18}$ cm$^{-3}$
\cite{Analytis}. Therefore, all carrier densities are fixed and the only
parameters that can be tuned to reproduce $R_{S}(0)$ and $R_{yx}(B)$ are
the mobilities of 2DEG, Dirac, and bulk electrons. Figure S9 shows the
result of the best fitting, which actually does not fit the $R_{yx}(B)$
behavior at all and obviously the slope of the calculated $R_{yx}(B)$ is
way too steep. This is essentially because the concentrations of both
Dirac and bulk electrons are too low in the 2DEG scenario. As presented
in the main text, the alternative scenario to ascribe the SdH
oscillations to the topological SS and consider a slight upward band
bending (which precludes the formation of the 2DEG state) naturally
explains the relatively large bulk carrier density and gives a totally
consistent description of our data.

It is worth noting that the above-estimated $E_F$ of 82 meV for the 2DEG
state in the downward band-bending scenario is already significantly
smaller than that seen in the ARPES data shown in Fig. S8, where $E_F$
is about 200 meV for the 2DEG state. This means that in our 10-QL film,
even if one assumes that the 2DEG state were formed and were responsible
for the SdH oscillations, the accumulated charge on the surface must be
much smaller than in the case observed in ARPES experiments
\cite{Hofmann}, and hence the band bending, if existed, must be much
weaker \cite{Poisson}. In other words, the SdH data dictate that the
band bending as strong as that shown in Fig. S9 cannot actually be
conceived for our 10-QL film.

In addition, we note that we did not observe any sign of progressive
downward band bending in our transport data, even though our samples are
not protected by any passivation layer and we applied a couple of
different sample preparation procedures: some samples were transferred
into an inert atmosphere immediately after the growth and were exposed
to air for only several minutes before measurements, other samples spent
hours in air during the contact preparation, and several samples have
been remeasured months after their first measurement. So far no
relaxations in transport properties have been detected in our MBE films,
in contrast to what we had observed in cleaved bulk single-crystal samples
\cite{BSTS} or what has been reported in the literature \cite{Analytis,
Fisher_nm, ACSnano}. It is possible that our films, which are of
high-enough quality to show SdH oscillations, are inherently more stable
than films of lower quality; namely, if the selenium-terminated
surfaces have smaller number of defects and hence contain fewer dangling
bonds, their chemical stability is expected to be higher \cite{Oleg} and
they are better protected against oxidation \cite{ACSnano} and the
resulting charging effect.

\begin{flushleft} 
{\bf S6. Film quality across $t_c$}
\end{flushleft} 

To confirm that the observed diminishment of the surface transport in
ultrathin films is not caused by a lowering of the quality in thinner
films, we have characterized the morphology of our films using atomic
force microscopy. As shown in Fig. S10, our ultrathin films remain
essentially flat and smooth across $t_c$ (surface bumpiness of only
$\sim$1 QL). This observation suggests that it is not very likely that
there is a drastic degradation in the film quality as a function of $t$.

It is worth noting that in our films the sheet density of bulk carriers
[shown in Fig. 3(a) of the main text by circles] changes smoothly but it
is not proportional to $t$, which is similar to what was previously
reported \cite{Kim-Oh}; the observed trend suggests that the density of
Se vacancies gradually increases in thinner films \cite{Kim-Oh}.

\begin{flushleft} 
{\bf S7. Weak antilocalization}
\end{flushleft} 

The fitting parameters of the WAL behavior measured in our films are shown
in Table I. One can clearly see that both $\alpha$ and $L_{\phi}$ rapidly
decrease below the critical film thickness. A similar tendency has also
been observed in previous thickness-dependence studies
\cite{e-e,Kim-Oh,Oh1}, although the decrease of $\alpha$ was less
pronounced and more gradual, which is not a surprise for more metallic
films where the suppression of the topological SS is less crucial. 

We also measured the angle dependence of the WAL behavior. Figure S11
shows the low-field magnetoresistance measured in the 10-QL film in
tilted magnetic fields. The WAL response plotted as a
function of $B \cos \theta$ clearly demonstrates its entirely 2D
character and is well fit by the Hikami-Larkin-Nagaoka (HLN) formula
[Eq. (1) of the main text].

\begin{table}
\centering 
\begin{tabular}{c c c} 
\hline\hline 
Thickness (QL) & $\alpha$ & $L_{\phi}$ (nm)  \\ 
[0.5ex] 
\hline 
2 & 0.007 & 101  \\ 
3 & 0.04 & 153  \\ 
5 & 0.5 & 256  \\ 
8 & 0.43 & 409  \\ 
9 & 0.45 & 584  \\ 
10 & 0.5 & 578  \\ 
15 & 0.45 & 573  \\ 
21 & 0.5 & 794  \\ 
32 & 0.4 & 405  \\ 
50 & 0.57 & 905 \\ 
[1ex] 
\hline 
\end{tabular}
\caption{Parameters of the WAL behavior described by the 
HLN formula.} 
\end{table}


\begin{figure}[b]
\begin{center}
\includegraphics[height=6cm]{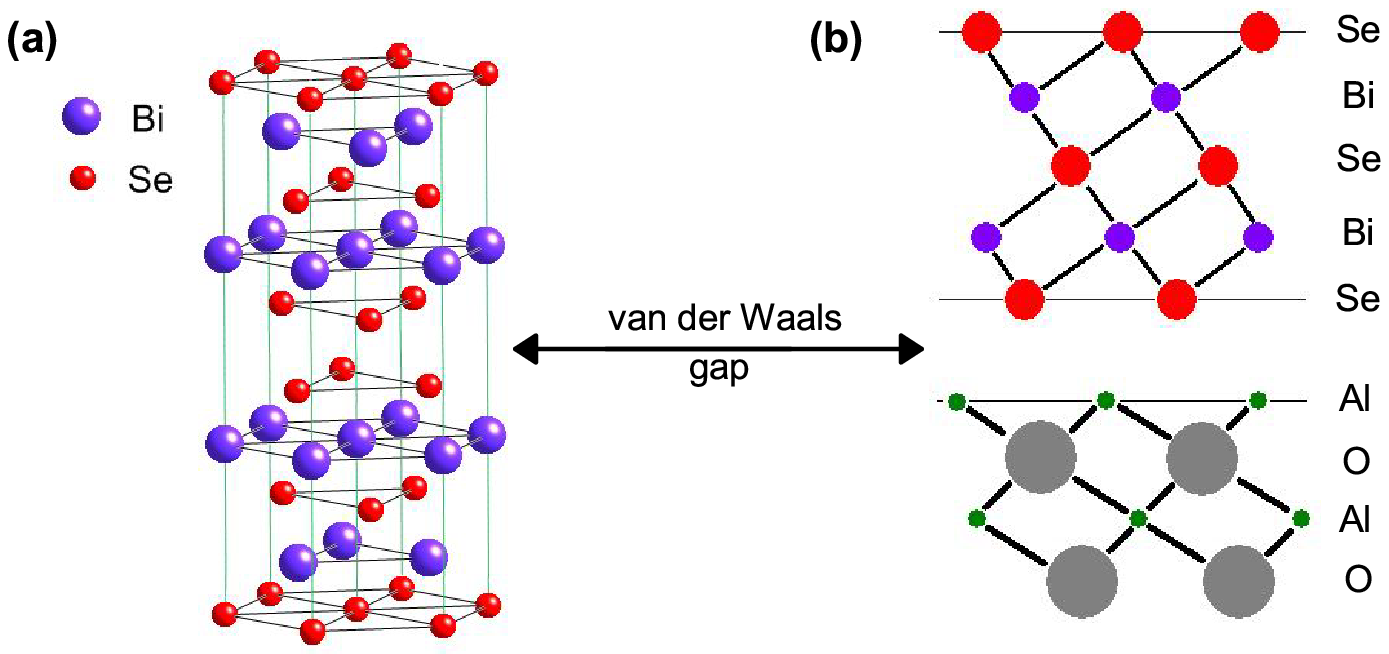}
\caption{
(a) Schematic picture of the Bi$_{2}$Se$_{3}$ crystal structure 
(two quintuple layers are shown).
(b) Schematic picture of the van der Waals MBE growth on sapphire.
} 
\end{center}
\end{figure}

\begin{figure}
\begin{center}
\includegraphics[height=8cm]{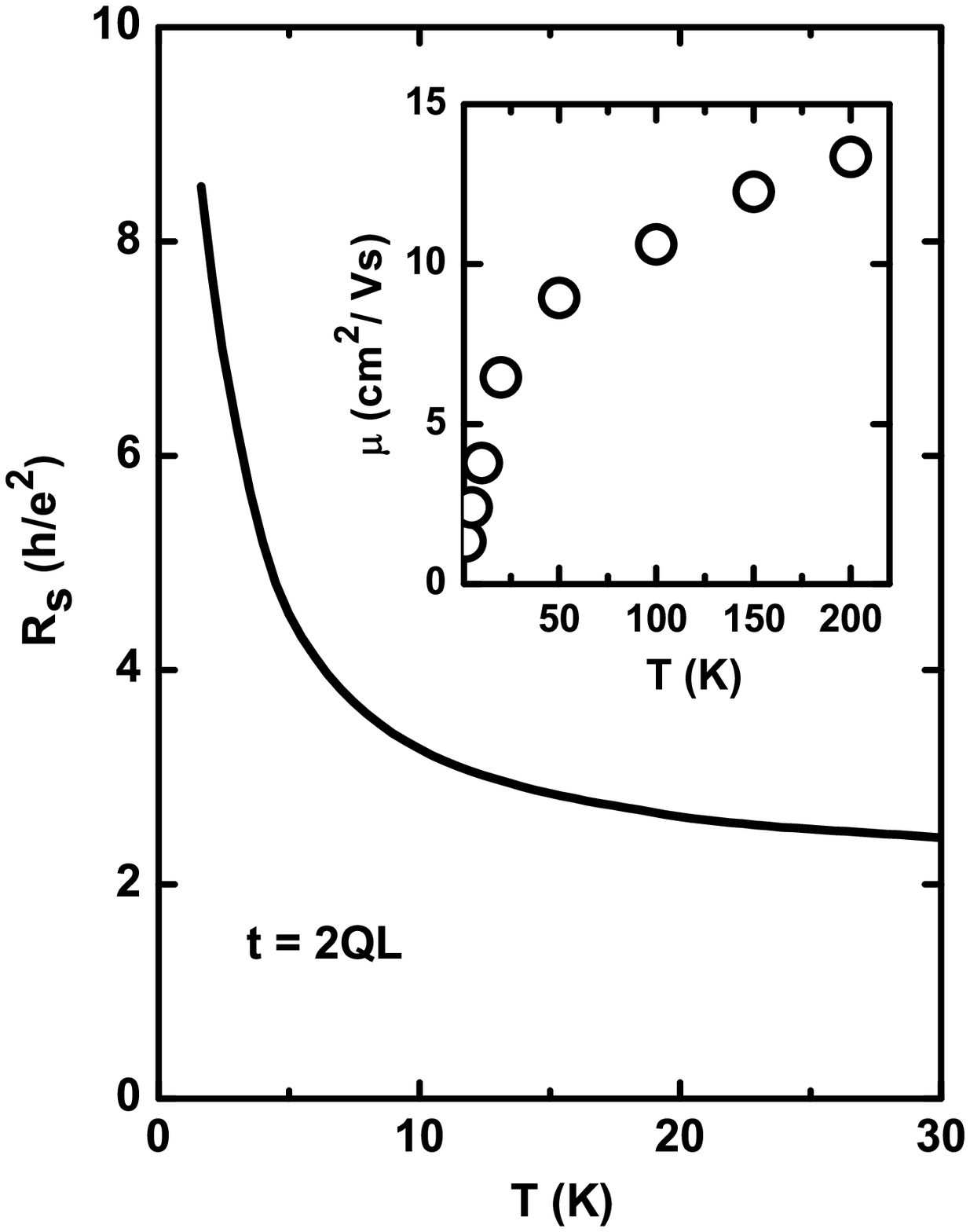}
\caption{
Temperature dependence of the sheet resistance in a 2-QL film showing 
a diverging trend at low temperatures. Inset shows the temperature dependence
of the Hall mobility $\mu = 1/enR_{s}$.
} 
\end{center}
\end{figure}

\begin{figure}
\begin{center}
\includegraphics[width=16.5cm]{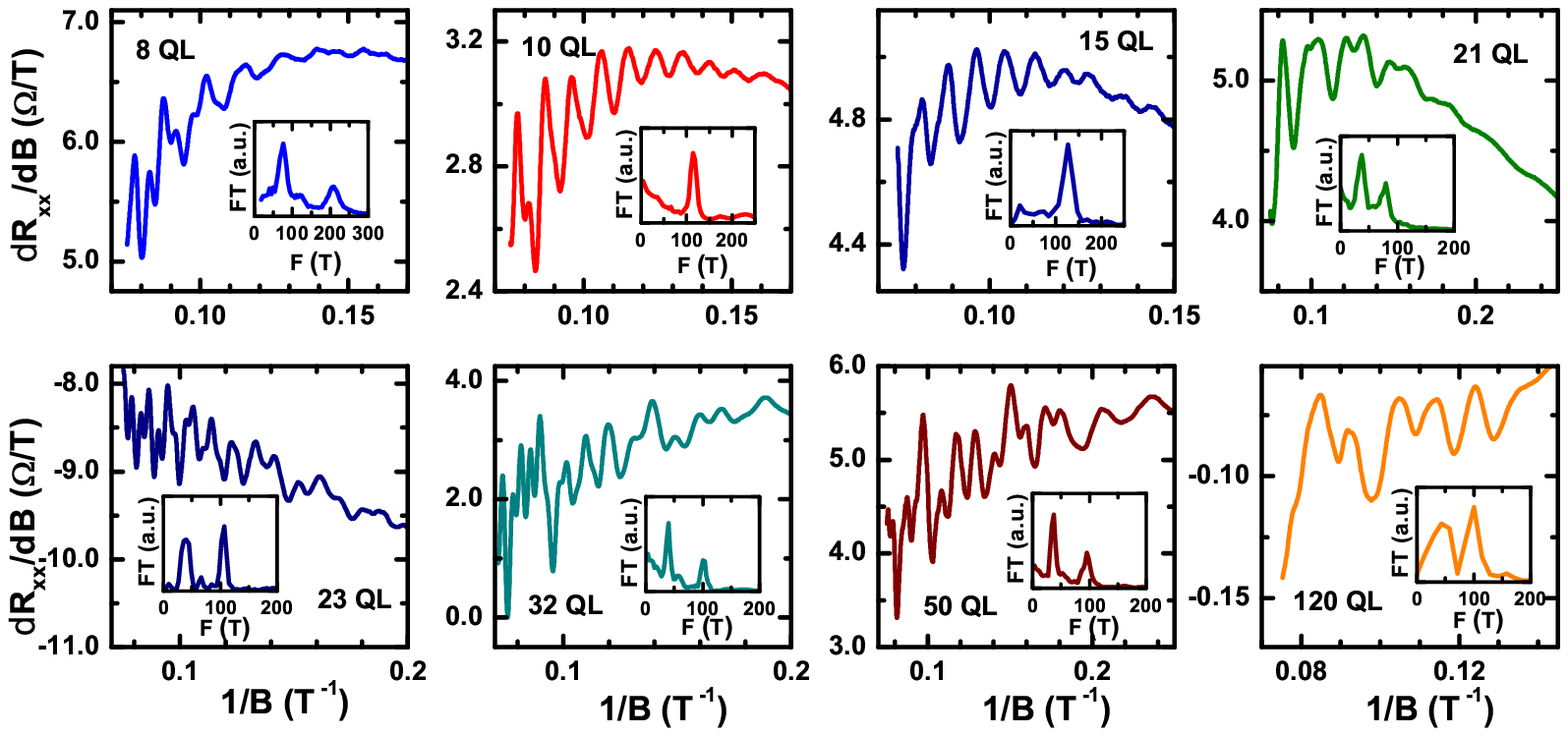}
\caption{
SdH oscillations in $dR_{xx}/dB$ measured in films with various 
thicknesses. Insets show the Fourier transform of the data in the main panel.
}
\end{center}
\end{figure}

\begin{figure}
\begin{center}
\includegraphics[height=7cm]{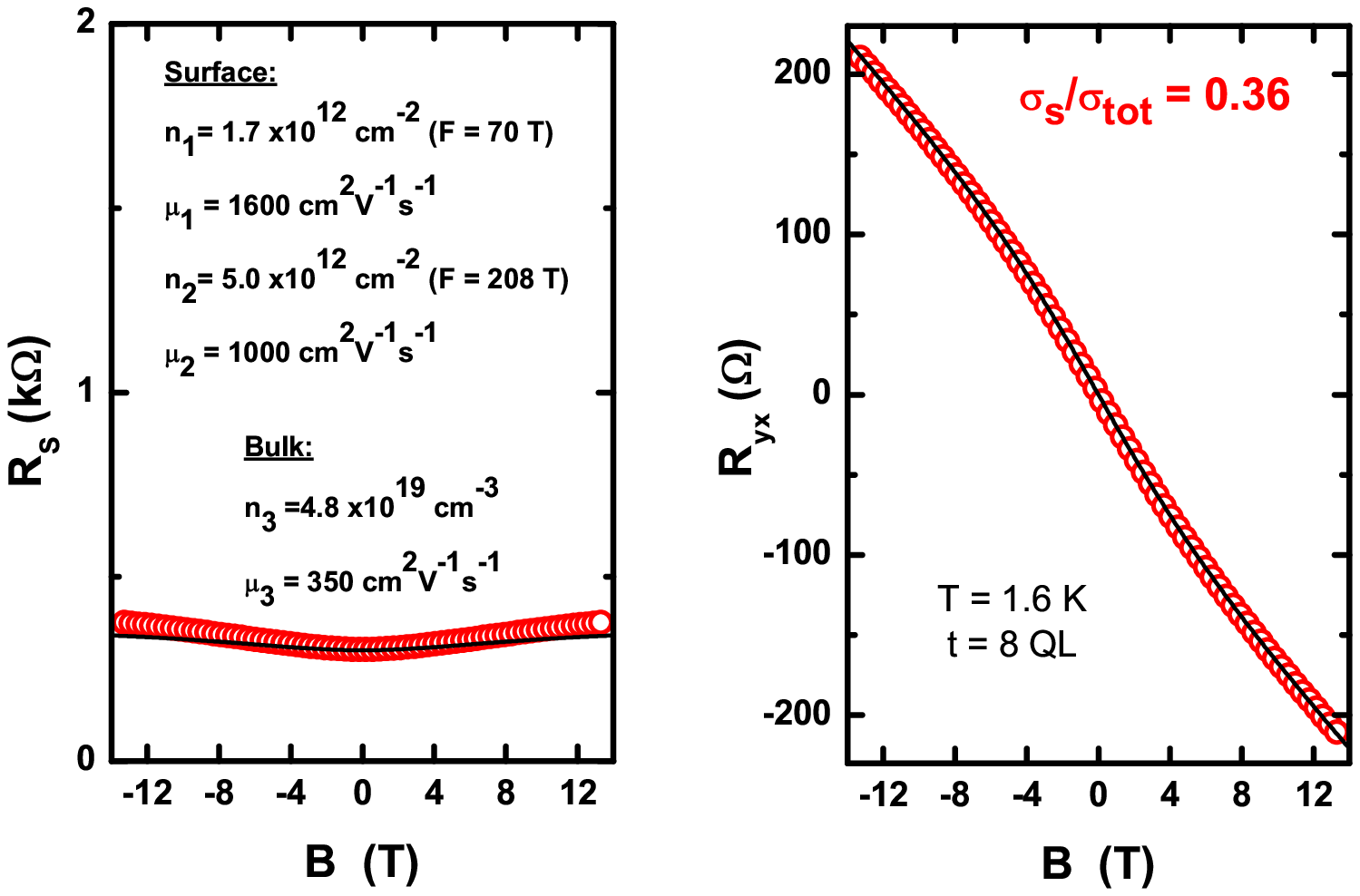}
\caption{
Fitting of a three-band model to $R_{S}(B)$ (left panel) and 
$R_{yx}(B)$ (right panel) measured in the 8-QL film at 1.6 K.
Symbols are experimental data. Solid lines are calculations.
} 
\end{center}
\end{figure}

\begin{figure}
\begin{center}
\includegraphics[height=7cm]{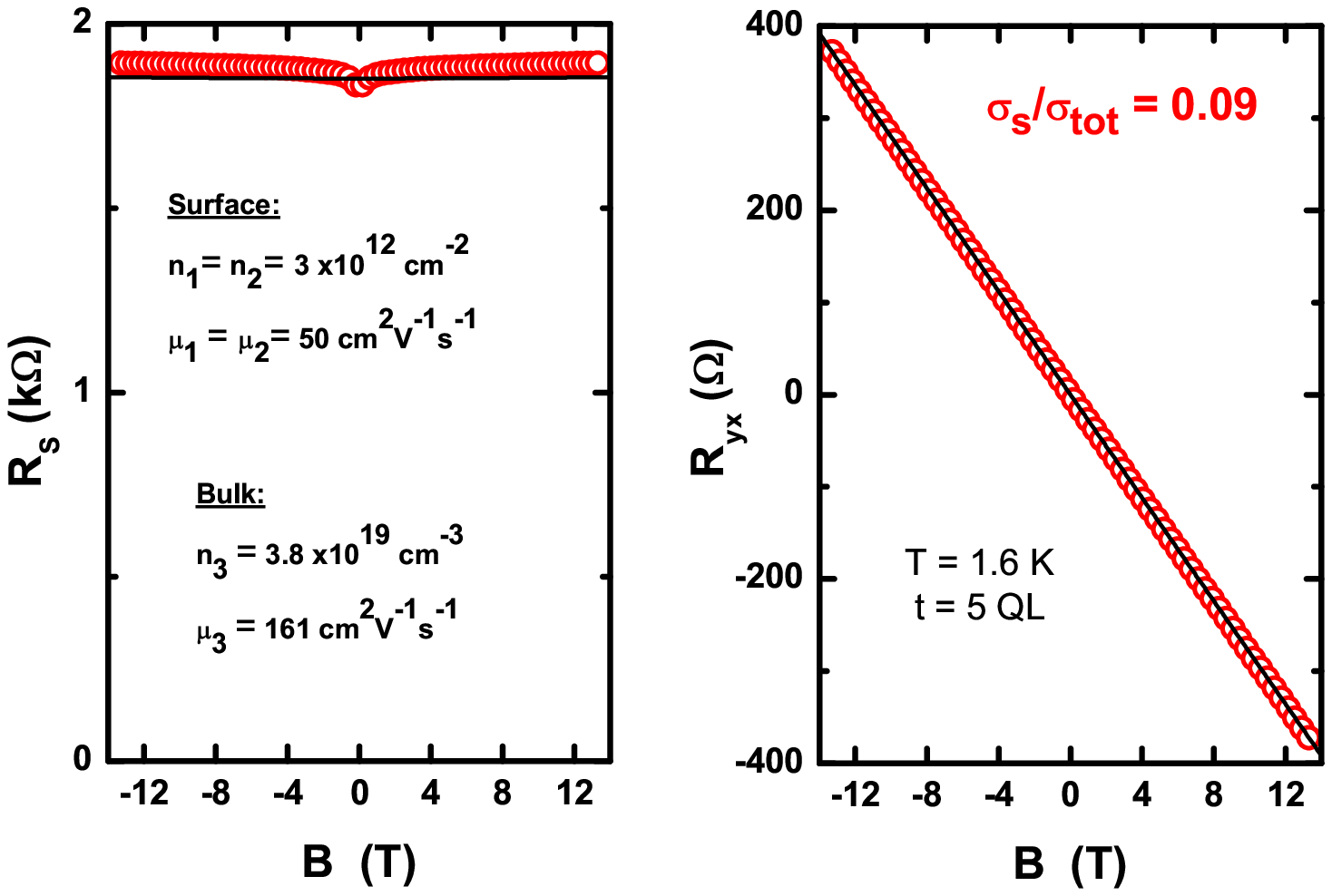}
\caption{
Two-band model calculations of $R_{S}(B)$ (left panel) and 
$R_{yx}(B)$ (right panel) for the 5-QL film with the parameters shown in
the left panel, together with the experimental data measured at 1.6 K.
} 
\end{center}
\end{figure}

\begin{figure}
\begin{center}
\includegraphics[height=7cm]{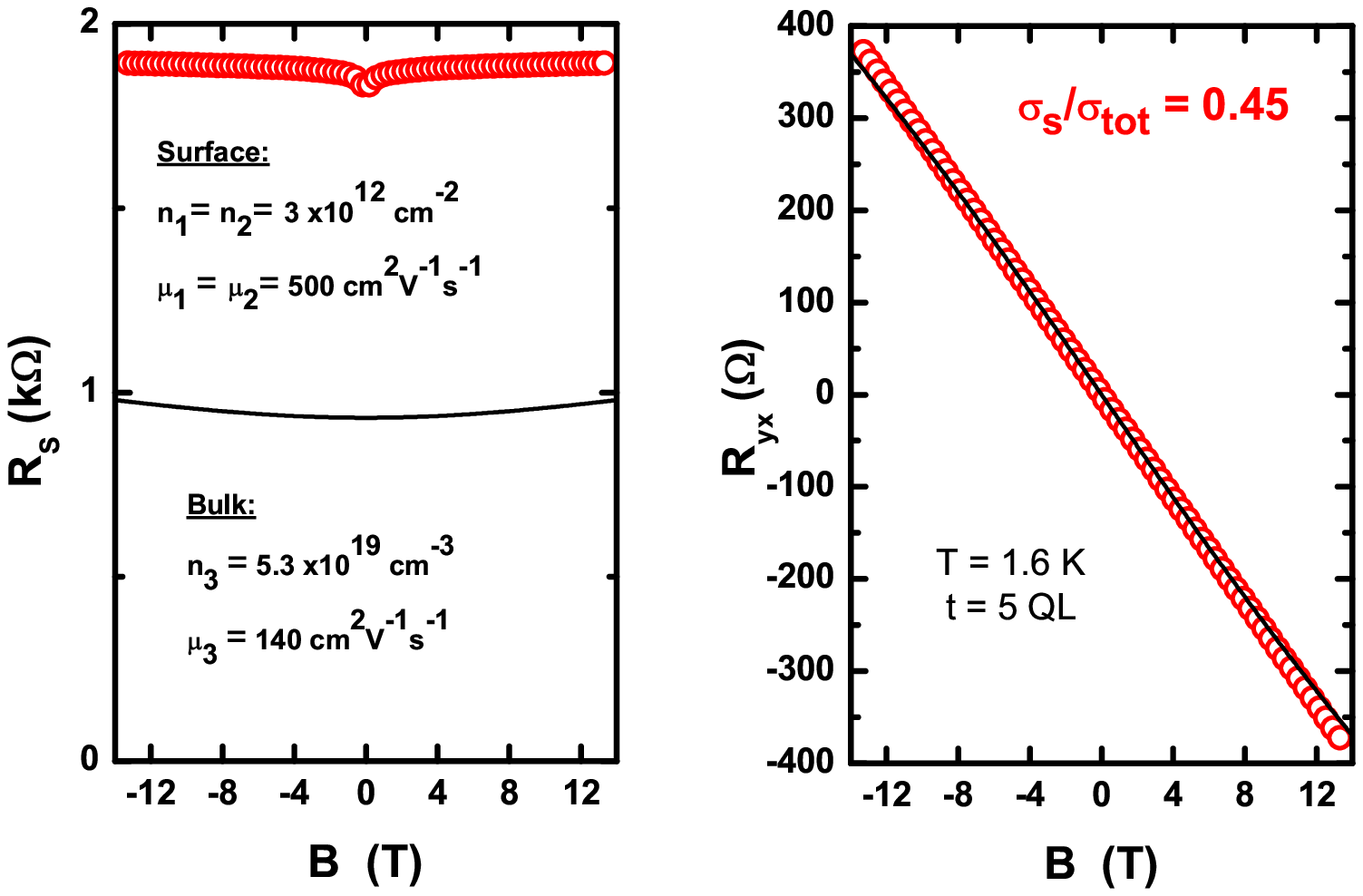}
\caption{
Two-band model calculations of $R_{S}(B)$ (left panel) and 
$R_{yx}(B)$ (right panel) for the 5-QL film with a hypothetical surface
mobility of 500 cm$^{2}$V$^{-1}$s$^{-1}$ which is lower than that for 
$t > t_c$ but is still reasonably high. The experimental data measured 
at 1.6 K are shown by symbols for comparison.
} 
\end{center}
\end{figure}

\begin{figure}
\begin{center}
\includegraphics[height=6cm]{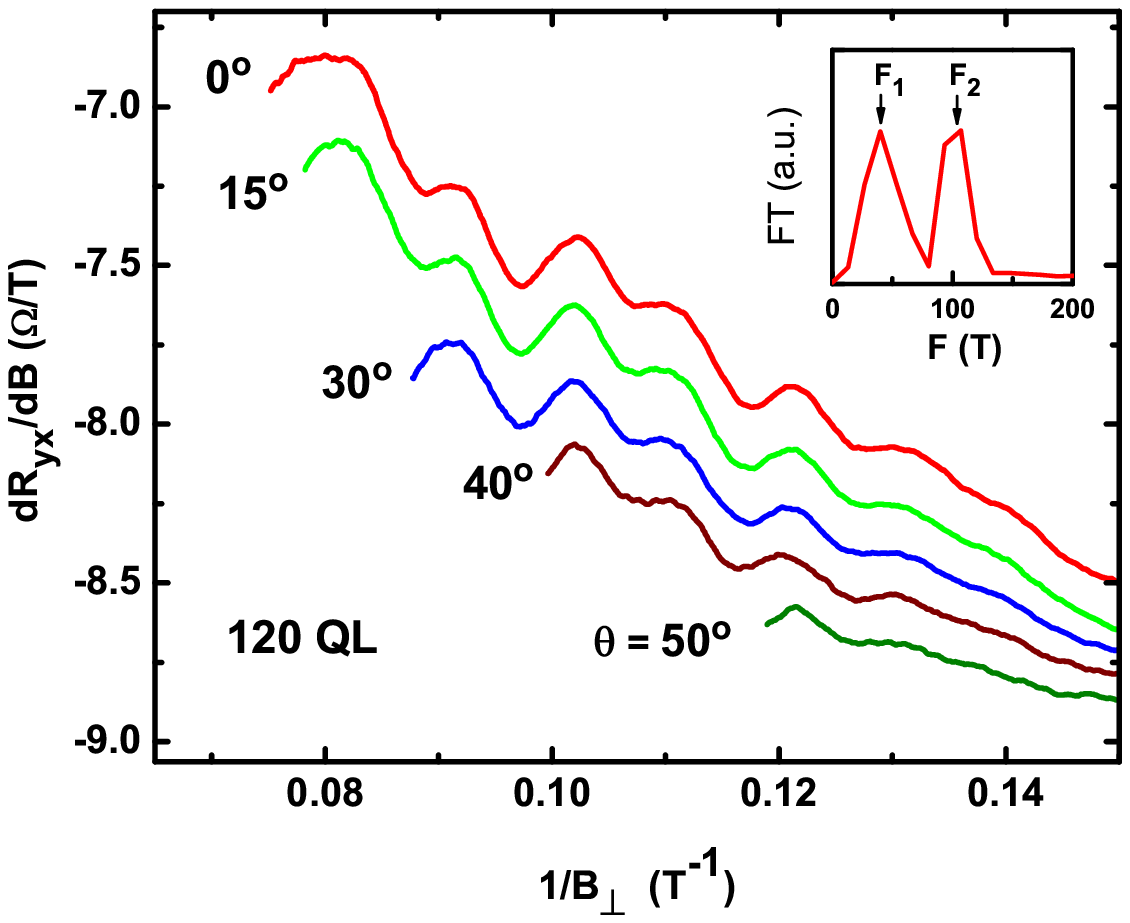}
\caption{
$dR_{yx}/dB$ (shifted for clarity) measured in tilted magnetic
fields in the Bi$_{2}$Se$_{3}$ film with the thickness of 120 QL and
plotted as a function of $1/B\cos\theta$, demonstrating the 2D character
of oscillations with two frequencies. Inset shows the Fourier transform
of the data at $\theta$ = 0$^{\circ}$ indicating the two-frequency nature 
of the oscillations.
}
\end{center}
\end{figure}

\begin{figure}
\begin{center}
\includegraphics[height=7cm]{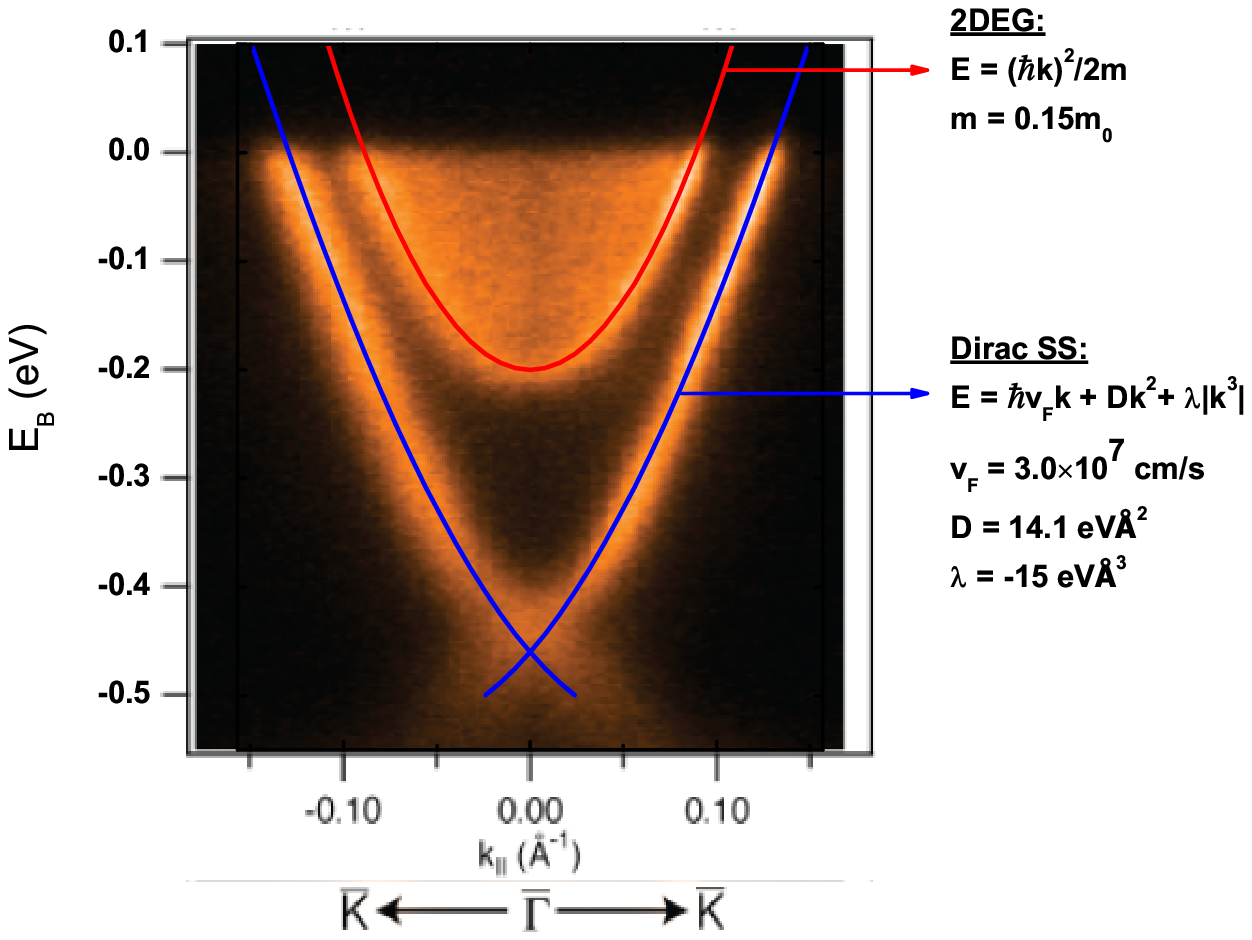}
\caption{
Energy dispersions of the SS and the 2DEG state measured on an aged 
surface of a cleaved Bi$_{2}$Se$_{3}$ crystal in Ref. \cite{Hofmann}.
Solid lines are our fitting results.
}
\end{center}
\end{figure}

\begin{figure}
\begin{center}
\includegraphics[height=7cm]{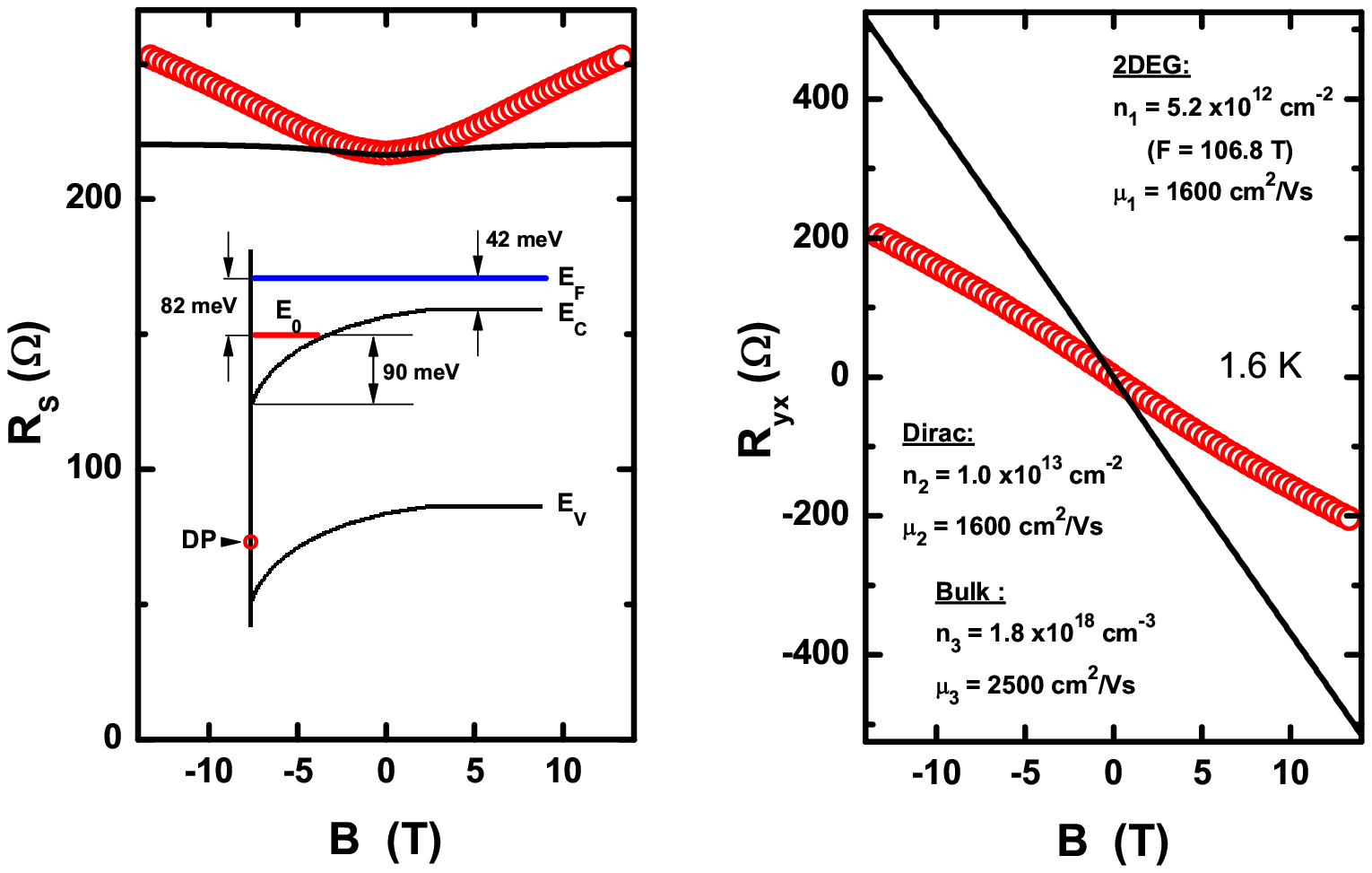}
\caption{
$R_{S}(B)$ and $R_{yx}(B)$ measured in the 10-QL film (symbols) and their 
fitting (solid lines) using parameters constrained in a 2DEG scenario 
(see text). Inset in the left panel shows a schematic picture of the band 
bending considered for the calculations, 
where $E_{C}$ is the bottom of the conduction band, $E_{V}$ is the 
top of the valence band, $E_{F}$ is the Fermi level, $E_{0}$ is the main 
(first) 2DEG energy state appeared due to the confinement of the electron 
motion perpendicular to the surface of the film.
}
\end{center}
\end{figure}

\begin{figure}
\begin{center}
\includegraphics[height=5.5cm]{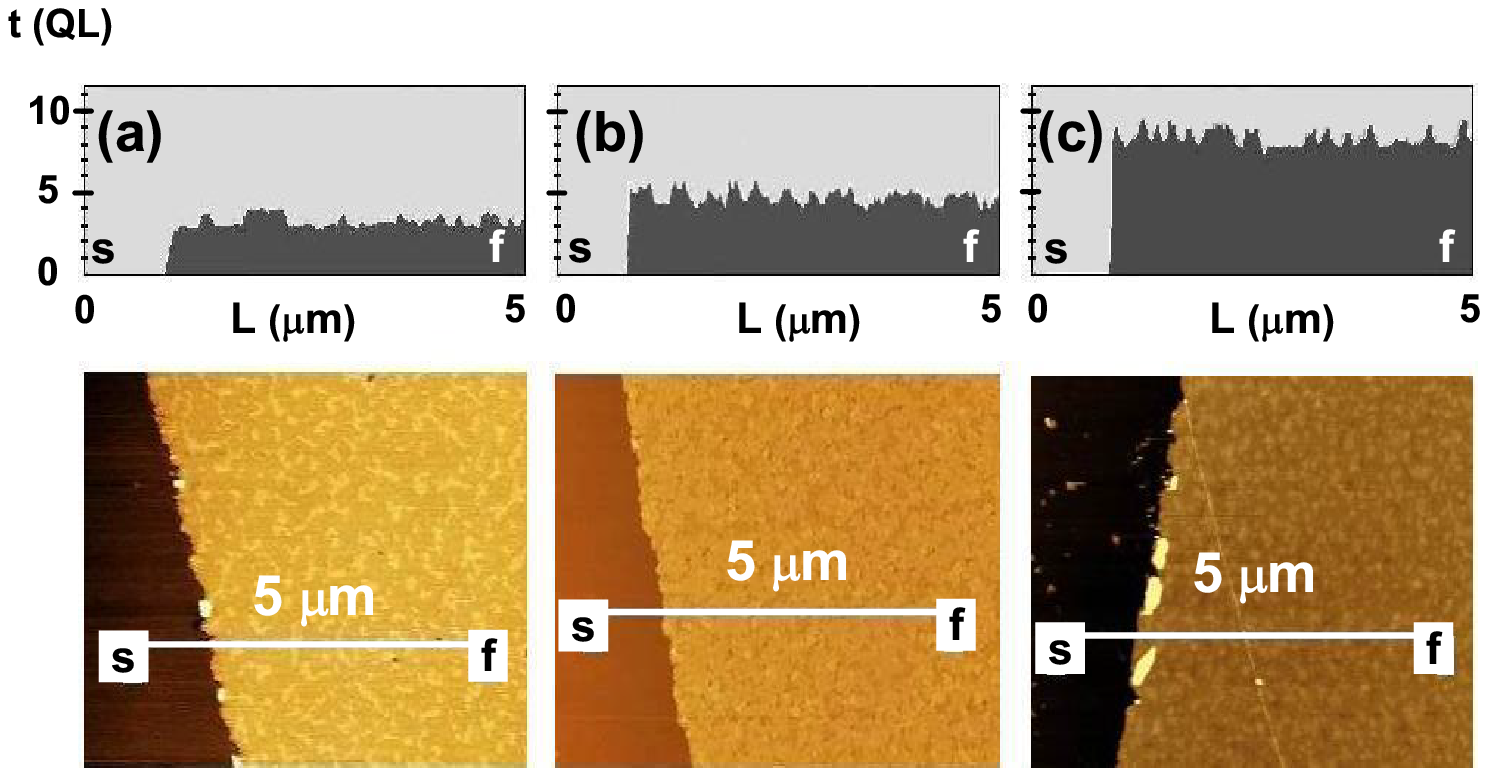}
\caption{
(a-c) Atomic force microscope images and height profiles of the 3, 5, 
and 8 QL films; the white lines indicate the 5-$\mu$m segment for which 
the profiles are shown.
}
\end{center}
\end{figure}

\begin{figure}
\begin{center}
\includegraphics[height=7cm]{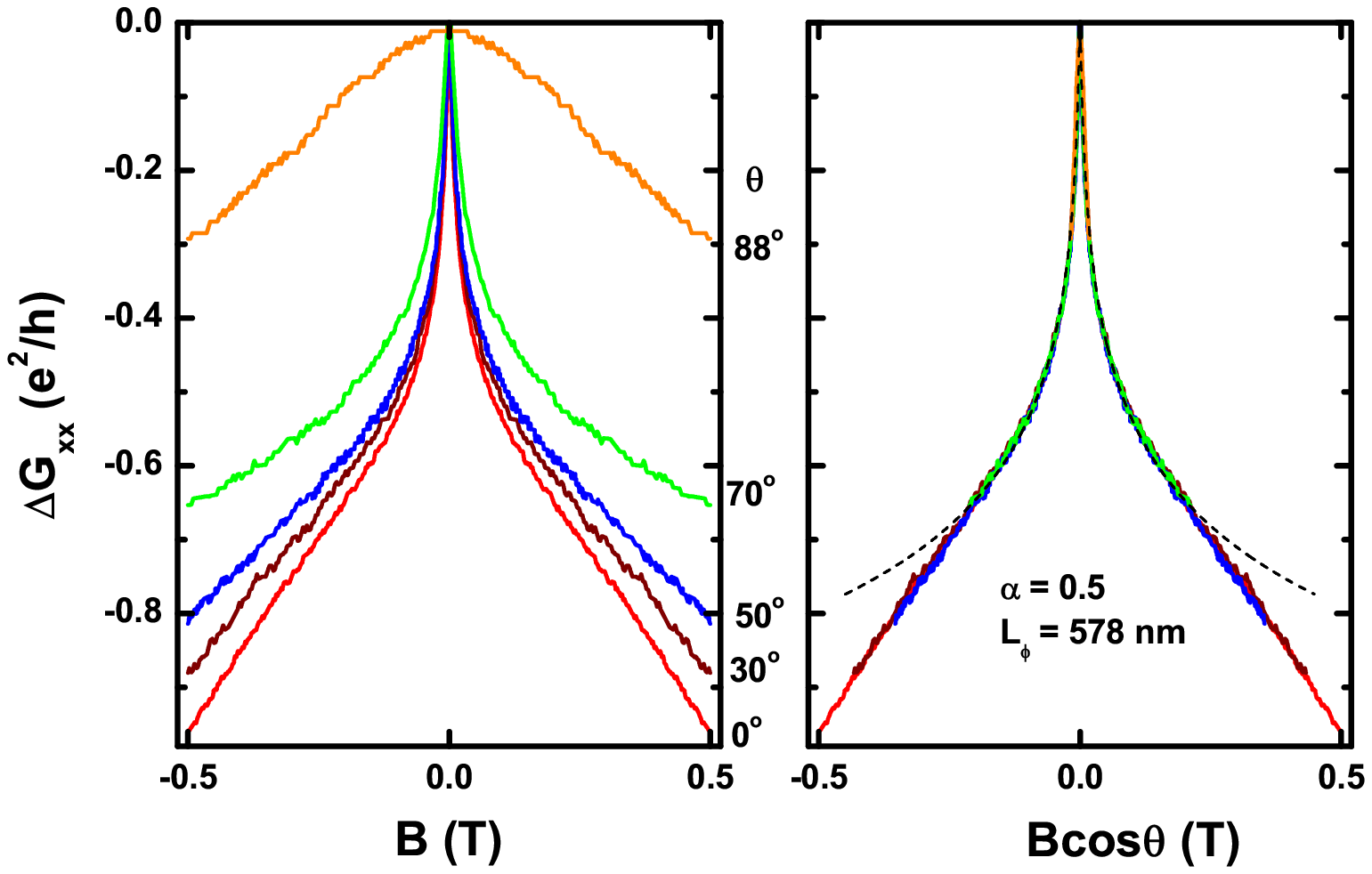}
\caption{
The WAL behavior measured for various magnetic field directions in the 
10-QL film. $\theta$ is the angle between the 
magnetic field and the surface normal. The dashed line is the fitting
of the HLN formula.
}
\end{center}
\end{figure}


\begin{thebibliography}{99}
\bibitem{footnote0}
Two-step deposition growth has been applied to various systems in the past.
For Bi$_{2}$Se$_{3}$, it has been recently reported in Refs. \cite{Xie, Oh}.
\end{thebibliography}
\end{document}